\documentclass[useAMS,usenatbib,usegraphicx]{mn2e}
\usepackage{times}
\usepackage{amssymb}
\usepackage{graphicx}
\usepackage{longtable}
\usepackage{subfig}

\newcommand{\kms}{$\rm km\,s^{-1}$}
\newcommand{\am}{$^{\prime}$}

\newcommand{\hi}{\mbox{H\,\sc{i}}}
\newcommand{\hii}{\mbox{H\,\sc{ii}}}

\newcommand{\gone}{G\,126.1--0.8--14}
\newcommand{\sixty}{$60\,\mu \rm m$}

\newcommand{\g}{G\,126.1--0.8--14}

\title[Triggered star formation in a molecular shell created by a SNR? ]{Triggered star formation in a molecular shell created by a SNR? }

\author[S. Cichowolski et al.] {S. Cichowolski$^1$\thanks{Member of
    the Carrera del Investigador Cient\'{\i}fico of CONICET,
    Argentina.}, S. Pineault$^2$, R. Gamen$^{3,4 *}$, E.M. Arnal$^{3,5 *}$,
  L.A. Suad$^5$\thanks{Post-Doc Fellow of CONICET, Argentina}, and M. E. Ortega$^{1 *}$\\
$^{1}$ Instituto de Astronom\'{\i}a y F\'{\i}sica del Espacio (IAFE), CC 67, Suc. 28, 1428 Buenos Aires, Argentina\\
$^{2}$ D\'epartement de physique, de g\'enie physique et d'optique, Universit\'e Laval, Qu\'ebec, G1V 0A6 Canada, and Centre de recherche en astrophysique du Qu\'ebec (CRAQ)\\
$^{3}$ Facultad de Ciencias  Astron\'omicas y Geof\'{\i}sicas, Universidad Nacional de La Plata, Paseo del
  Bosque s/n, 1900 La Plata, Argentina\\
$^{4}$ Instituto de Astrof\'{\i}sica de La Plata (CCT-La Plata, CONICET), Observatorio Astron\'omico
Paseo del Bosque s/n, La Plata, Argentina\\
 $^{5}$ Instituto Argentino de Radioastronom\'{\i}a  (CCT-La Plata, CONICET), CC 5, 1894 Villa Elisa,
  Argentina}

\begin{document}

\date{}
\pagerange{\pageref{firstpage}--\pageref{lastpage}} \pubyear{2002}

\maketitle

\label{firstpage}

\begin{abstract}

We present a study of a new molecular shell, \g, using available multiwavelegth Galactic plane surveys and optical Gemini observations. A well defined shell-like structure is observed in the CO(1--0) line emission at ({\it l,b}) = (126\fdg1, --0\fdg8), in the velocity range --10.5 to --15.5 \kms.
The \hi\, emission shows a region of low emissivity inside \g, while radio continuum observations reveal faint non-thermal emission possibly related to this shell.
Optical spectra obtained with Gemini South show the existence of B-type 
stars likely to be associated with \g. An estimate of the stellar wind
energy injected by these stars show that they alone can not be able to create such a structure. On the other hand, one supernova explosion would provide enough energy to generate the shell. 
Using the {\rm MSX}, {\rm IRAS}, and {\rm WISE} Point Source Catalogues we
have  found about 30 young stellar objects candidates, whose birth could have been triggered by the expansion of \g. In this context, Sh2-187 could be a consequence of the action on its surroundings of the most massive (and thus most evolve) of the stars formed by the expanding molecular shell.

\end{abstract}


\section{Introduction}

Massive stars have strong winds and radiation fields which produce ionized regions while clear out parsec-scale cavities in the interstellar medium (ISM) \citep{cas75,wea77}. At the boundaries of these cavities lies the material displaced by the shock fronts, forming dense shell-like structures. Such a dynamic process strongly modifies the structure and dynamics of the molecular clouds and may trigger the formation of new stars \citep{elm98}.

It is in this modified environment where the massive star ends its life exploding as a supernova and injecting about $10^{51}$ erg of kinetic energy into the ISM. The ejected material drives a blast wave into the ISM producing a supernova remnant (SNR).
There are about 230 known SNRs in the Galaxy, most of which have been discovered in radio continuum and/or X-rays \citep{gre09}. This number is much less than what we would expect ($>1000$, \citet{li91}) from the Galactic supernova rate and their life time.
This deficit is likely the result of selection effects acting against the discovery of old ($>10^5$ yr), faint remnants.
Although the detection of expanding \hi\, shells is difficult as most of the known SNRs are located in the Galactic plane, where the Galactic background \hi\, emission causes severe contamination, 
looking for shell structures in the atomic and molecular gas could be a good approach to detect the oldest SNRs.

In this paper we present a study of a new molecular shell-like structure, which we called \g.
Based on optical, radio continuum, infrared, molecular, and atomic data we carried out a multi-wavelength study of the region trying to disentangle the origin of the molecular structure, its possible relation to Sh2-187, and its eventual role in the process of triggering new stars.

\section{Observations}

\subsection{Radio Line and Continuum}

Radio continuum data at 408 and 1420 MHz, as well as 21 cm spectral
line data, were obtained using the Dominion Radio Astrophysical Observatory (DRAO) interferometer as part  of the Canadian Galactic Plane Survey (CGPS, \citet{tay03}). 
As a result of the
observations, the CGPS provides a 256 velocity channel data-cube of
the \hi\, spatial distribution together with 1420 MHz and 408 MHz
continuum images. 
Molecular CO $J= 1-0$ observations were obtained from the Five College Radio Astronomical Observatory
(FCRAO) CO Survey of the Outer Galaxy \citep{hey98}.
The observational parameters are listed in  Table \ref{tabobsparams}.
The angular resolution of the radio  continuum and \hi\, line images  varies slightly over the field covered by \g\,  and we simply took the mean.

\subsection{Infrared}

At infrared wavelengths, we have used the IRAS high-resolution (HIRES) data \citep{fow94} and data retrieved from the Midcourse Space Experiment (MSX) Galactic Plane Survey \citep{pri01}. Table \ref{tabobsparams} summarizes some of the relevant observational parameters.  
For the \sixty\ image, the resolution is somewhat
more variable than in the radio images and then we took the median value.
 
The IRAS Point Source Catalog v2.1 (PSC, \citet{bei88}), the Midcourse Space Experiment (MSXC6, \citet{ega03}), the 2MASS All-Sky Point Source Catalog  \citep{skr06}, and the WISE All-Sky Source Catalog \citep{wri10} were also used in this work.

\subsection{Optical}
Optical spectra of four early-type stars seen in projection against the field of the shell were obtained with the GMOS at Gemini Observatory under the Poor-weather proposal GN-2012B-Q-134 (PI: SC). We employed the GMOS in its long-slit mode, with the B600 grating and a slit-width of 0.5 arcsec, which
resulted in a wavelength range coverage between 3780\AA\, and 6700\AA\, and a resolution R $\sim$ 1700.

\begin{table}
\caption{Observational parameters}\label{tabobsparams}
\vskip 0.25truecm
\centering
\begin{tabular}{l c c }
\hline\hline
Band & Synthesized beam$^a$   & rms noise$^b$ \\
\hline
1420 MHz   &  0\farcm92  $\times$  0\farcm82 $\quad$ 81.6 & 0.056 K \\
408 MHz    &  3\farcm15  $\times$  2\farcm82 $\quad$ 84.8 & 0.81 K \\
60 $\mu$m  &  1\farcm65  $\times$  0\farcm92 $\quad$ 128  & 0.38 MJy/sr \\
8.3 $\mu$m & 18\farcs4 & $  10^{-7}$ W\, m$^{-2}$\,sr$^{-1}$ \\
HI & 1\farcm1  $\times$  1\farcm0 $\quad$ 81.6 & 2.3 K \\
CO  &45\farcs0 & 0.2 K  \\
\hline
\end{tabular}
\begin{list}{}{}
 \item {$^a$} Semi-major and semi-minor axes, and position angle,
   in degrees, measured counter-clockwise from the
   horizontal (longitude) axis.
 \item {$^b$} Noise for original full resolution images.
\end{list}
\end{table}

\section{CO emission distribution}\label{co}

\begin{figure*}
\includegraphics[width=15cm]{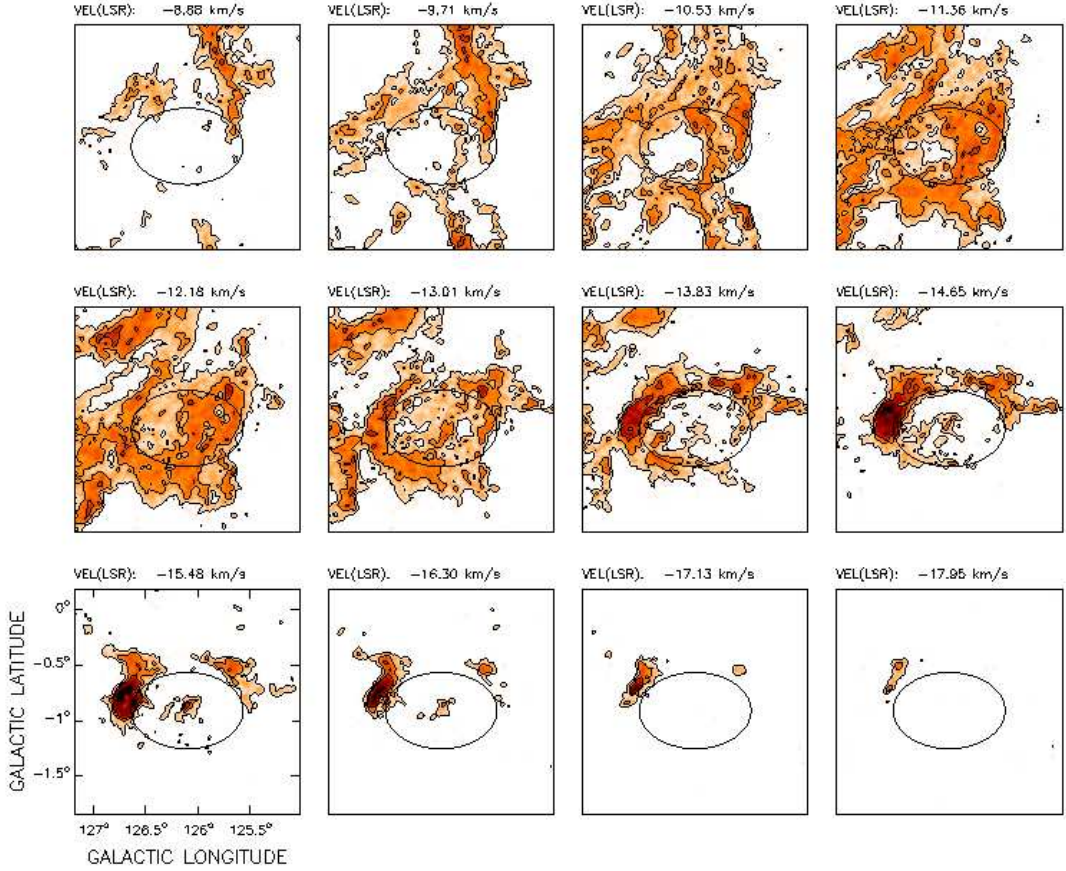}
\caption{ Channel maps of the CO(1-0) emission distribution between --9 and --18 \kms. The central velocity of each panel is given in the upper left-hand corner. Contours are at 1, 3, and 5 K. The ellipse shows the outline of \g. }
\label{set-co}
\end{figure*}

Figure \ref{set-co} shows a set of images of the CO(1-0) emission distribution within the velocity range from about --9 to --18 \kms\, (all velocities
are with respect to the Local Standard of Rest (LSR)).
A shell-like structure is clearly observed from --10.5  to about --15.5 \kms\,
   centered at ({\it l, b})  = (126\fdg1. --0\fdg8) (delineated by the ellipse in Fig. \ref{set-co}).
Since the structure is better defined at  about --14 \kms\,, 
 from here on we will refer to it as \g. 

 From Fig. \ref{set-co} it can be seen that there is no emission in the centre of the shell
that could be interpreted as the approaching and receding caps. This absence may
be indicating either that the structure has a ring morphology or, as pointed out
by \citet{caz05}, that there is significant velocity dispersion. The latter may result in a shell whose observed ring diameter appears not to vary significantly  from channel to
channel and whose receding and/or expanding caps become hard to detect.

\begin{figure*}
\includegraphics[width=13.0cm]{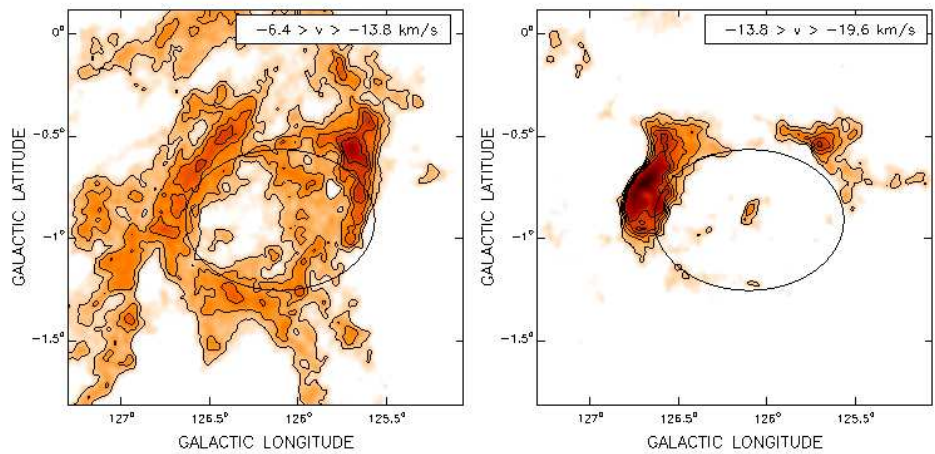}
\caption{CO emission distribution averaged in the velocity range from   --6.4 to --13.8 \kms\, (left panel) and from --13.8 to --19.6 \kms\, (right panel) showing the  receding and approaching halves of the shell. Contours are at 1, 1.5, 2, 2.5 and 3 K. The ellipse shows the outline of \g.}
\label{prom}
\end{figure*}

An almost circular strong molecular feature centered at ({\it l, b}) = (126\fdg68, --0\fdg82) is noticeable along the velocity range from about --14 to --16 \kms. 
This molecular structure is related to the \hii\, region Sh2-187,
which is a small (diameter about 9\am) and young ($\sim 2 \times 10^5$ yr) ionized region that was extensively studied by \citet{jon92}.
\citet{ros78} observed this ionized region in the H109$\alpha$ recombination line and derived a radial velocity of   V$_{H109\alpha} = -14.6 \pm 0.4$ \kms. This velocity is in agreement with the velocity range where the CO shell is better observed, allowing us to conclude that both features, Sh2-187 and \g\,, are likely at the same distance. 
In Fig. \ref{prom} we show the CO emission averaged in two velocity intervals, corresponding to the receding and approaching sides of the structure, if a systemic velocity of --14 \kms\, is assumed. It can be noticed that the molecular gas related to Sh2-187 is observed
  as part of the approaching half of \g. 

As part of a study of molecular clouds in the second Galactic quadrant, \citet{cas84} analyzed the region around $l = 126^{\circ}$ in the Orion arm and concluded that Sh2-187 belongs to a large molecular complex but, given  the small area observed, they do not recognize the molecular cloud as having  a shell morphology.
Later, \citet{yon97} carried out a $^{13}$CO(1--0) survey of the nearby molecular clouds in the region 100$^{\circ} < l < 130^{\circ}$ and --10$^{\circ} < b < 20^{\circ}$ and identified 188 distinct $^{13}$CO clouds. The one related to Sh2-187 was named 126.6-00.6, based on its peak position. \citet{yon97} derived for the cloud some parameters such as a size of 96\arcmin $\times$ 72\arcmin, a velocity of V$_{LSR} = -16.0$ \kms, a distance of 800 pc, and a molecular mass of 7600 M${\odot}$.

\citet{jon92} assumed a systemic velocity of --15.0 \kms\, for the molecular cloud related to Sh2-187, for which they inferred a kinematical distance of 1 kpc for the region. Based on spectroscopic and photometric observations of the candidate exciting stars of many ionized regions, \citet{rus07} inferred for the candidate exciting star of Sh2-187 a distance of 1.44 $\pm$ 0.26 kpc. From here-on we adopt this value as the distance of Sh2-187 and \g.

Several physical parameters can be derived from the $^{12}$CO data. 
As can be inferred from Fig. \ref{prom}, the angular size of the shell is about $\Delta\, l \times \Delta\,b =$ 1\fdg2 $\times$ 0\fdg9, or about 30  $\times$ 23 pc at the adopted distance. 
The mass of the shell can be obtained by integrating the CO line intensity as  $W{\rm_{CO} =\int{T(CO)dv}}$, where
T(CO) is the average temperature of the molecular gas over the considered
velocity interval. In this case we consider the velocity interval from $-8.9$ to --18 \kms.  To calculate the H$_2$ column density,
the  relationship $X=N({\rm H_2})/W_{{\rm CO}} = 1.9 \times 10^{20}\ {\rm
cm^{-2} (K\ km s ^{-1})^{-1}}$ \citep*{gre90,dig95} is used.  The molecular mass was derived from
$M[M_\odot]=4.2 \times 10^{-20}N(H_2)D^2 A$, where $D$ is the distance in pc and $A$  is the area in steradians.
 Taking into account all  the estimated values and their corresponding errors, we obtain $M_{\rm shell}[M_\odot]$ = (6.5 $\pm$ 3.1) $\times 10^4\, M_{\odot}$.
The kinetic energy stored in the shell can be estimated as $E_{\rm kin} = 0.5\, M_{\rm shell}\, V^2_{\rm exp}$, where $V_{\rm exp}$ is the expansion velocity of the shell. 
Adopting an expansion velocity equal to half the velocity interval 
where the structure is observed, $V_{\rm exp} = 4.5 \pm 0.8$ \kms\,,  and the mass estimated above we obtain 
$E_{\rm kin} =$ (1.3 $\pm$ 0.8) $\times 10^{49}$ erg.

\section{HI emission distribution}\label{hi}

\begin{figure*}
\includegraphics[width=15cm]{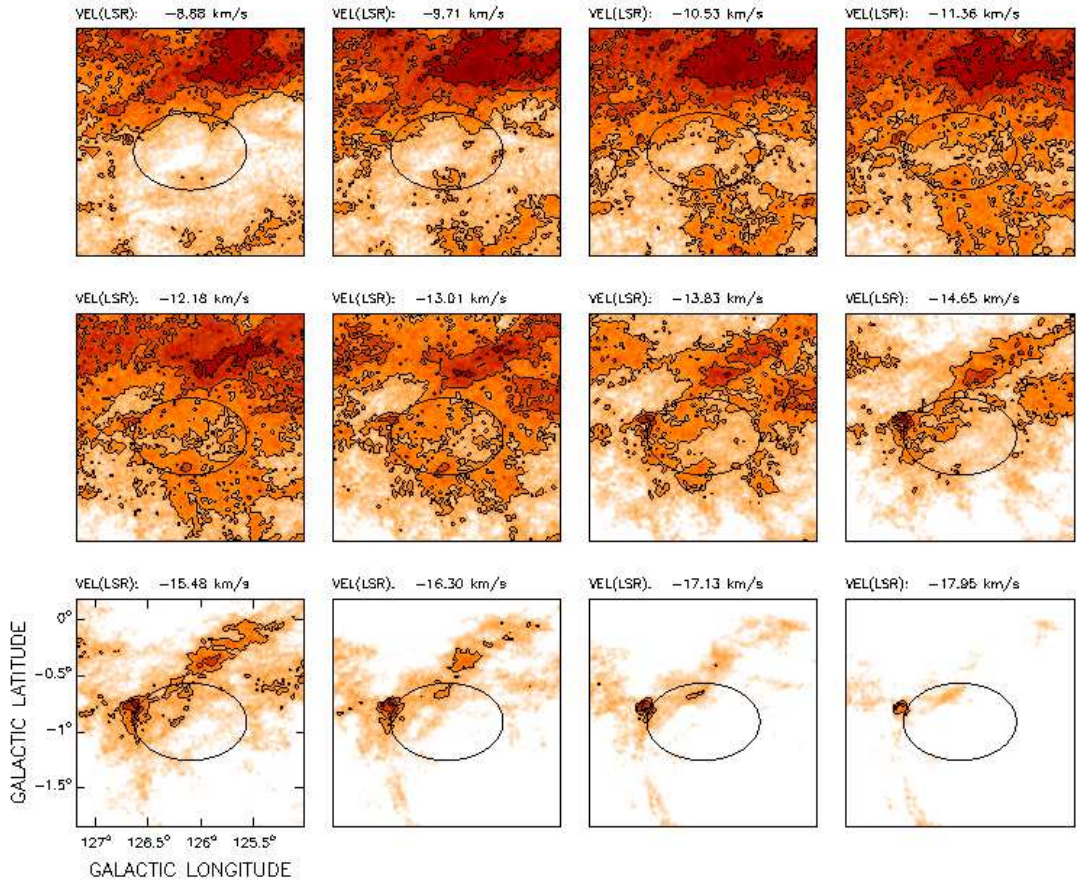}
\caption{ Channel maps of the HI emission distribution between --9 and --18 \kms. The central velocity of each panel is given in the upper left-hand corner. Contours are at 55, 65, and 75 K. Spatial resolution is 2\arcmin. As in Fig. \ref{set-co} the ellipse shows the outline of \g. }
\label{set-hi}
\end{figure*}

In order to analyze if the molecular shell structure has a counterpart in the atomic gas, we inspect the HI data cube obtained from the CGPS.
In Fig. \ref{set-hi} we present a set of images showing the HI emission distribution in the same velocity interval where the molecular emission from \g\, is observed. The images have been smoothed to a resolution of 2\arcmin.
Although the HI emission distribution shows a more complex  structure, a similar spatial and
kinematical distribution of the HI and molecular gas is clearly observed. As observed in the CO emission, the HI  shell is better defined at --14 \kms. Towards more negative velocities the major contribution comes from the gas related to Sh2-187.

Assuming the gas is optically thin, the excess \hi\, mass in the shell is given by $M_{\hi}$ ($M_{\odot}$) = $1.3 \times 10^{-3}\, D^2 ({\rm kpc})\, \Delta v$ (\kms) \, $\Delta T$ (K)\,$\Omega_{\rm sh}$ (am$^2$), where  $\Delta v$ is the velocity width over which the \hi\, shell is being detected,  $\Delta T$ is the difference between the average shell temperature and the background temperature, $\Delta T$ = ($T_{\rm sh} - T_{\rm bg}$), $D$ is the distance, and $\Omega_{\rm sh}$ is the angular size of the shell. 

\begin{figure}
\includegraphics[width=8cm]{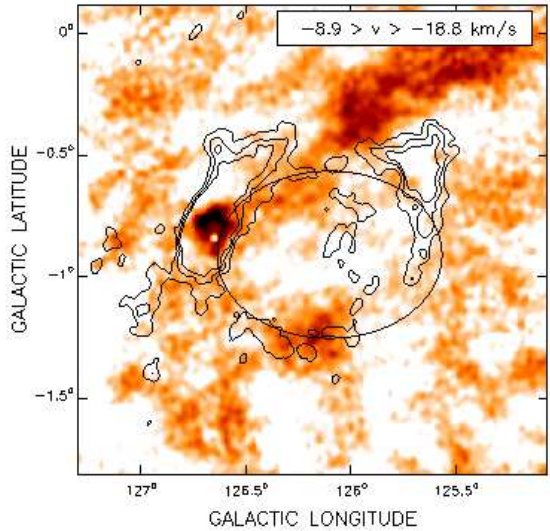}
\caption{Averaged \hi\, emission within the velocity interval --8.9 to --18.8 \kms\,, after removing the large scale structures. The contours correspond to the averaged CO emission at 1.5, 2, and 2.5 K. The ellipse indicates the location of \gone, as in previous figures. }
\label{hi-co}
\end{figure}

As can be seen from Fig. \ref{set-hi}, the determination of the angular size of the \hi\, shell is difficult due to the presence of a nonuniform background. So, in order to get rid of the large-scale structures, we smooth the original image to a 30\arcmin\, resolution and subtract it from the 2\arcmin\, resolution image. The resulting image is shown in Fig. \ref{hi-co}, where the CO emission related to the shell is also shown by the solid  contours levels. The morphological agreement between the atomic and molecular emissions is observed. 
Thus, taking $\Delta v = 10 \pm 2$ \kms, $\Delta T = 8 \pm 2$ K, $D = 1.44 \pm 0.26$ kpc, and $\Omega_{\rm sh} = 1500 \pm 300$ am$^2$, we obtain $M_{\hi} = 323 \pm 169 M_{\odot}$.

A  rough estimate of the kinematic age   of \g\, can be obtained using a simple model to describe the expansion of 
a shell created by an injection of mechanical energy, $t_{\rm dyn} = \alpha \, R /V_{\rm exp}$, where $\alpha = 0.25$ for a radiative SNR shell and $\alpha = 0.6$ for a shell created by the action of stellar winds \citep{wea77}.
Considering an effective radius of 13 pc, we obtain $t_{\rm dyn}$ = (0.7 $\pm$ 0.2) $\times 10^6$ yr ($\alpha = 0.25$) and $t_{\rm dyn}$ = (1.7 $\pm$ 0.5) $\times 10^6$ yr ($\alpha = 0.6$).

\section{Diffuse continuum emission}\label{rc}

\begin{figure*}
\includegraphics[width=16cm]{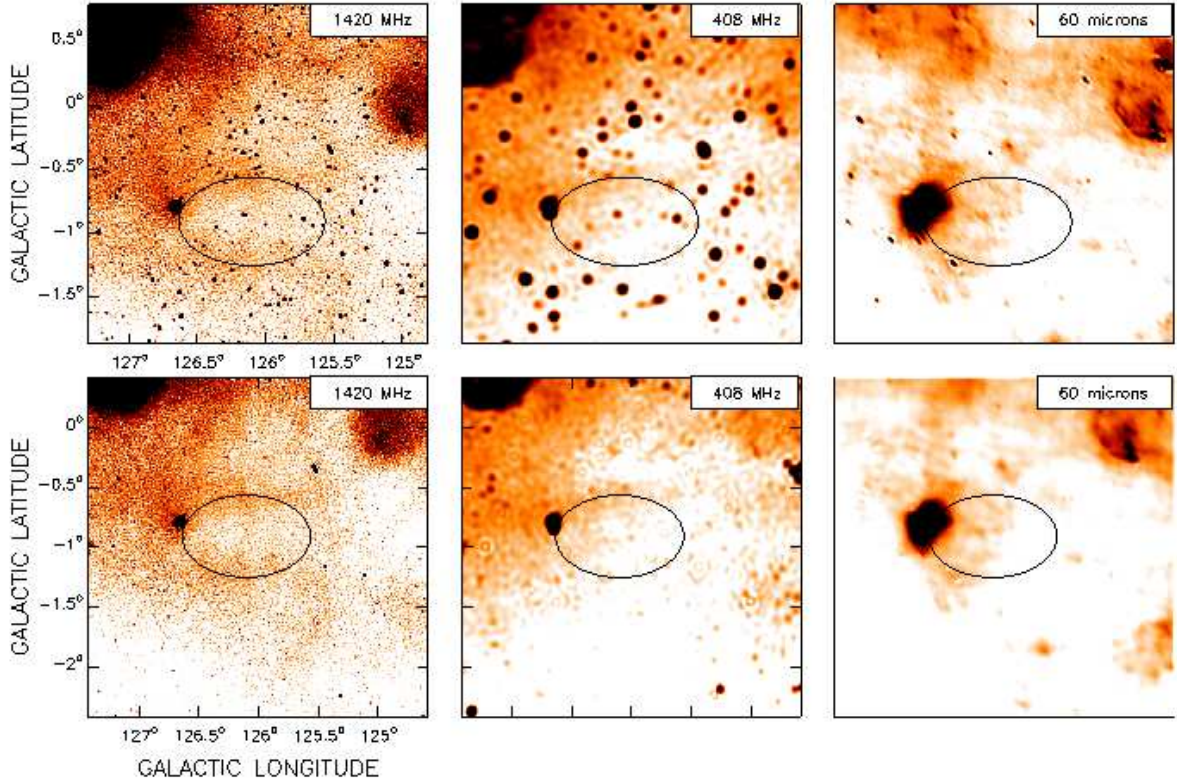}
\caption{ Radio continuum at 1420 MHz ({\it left panels}), 408 MHz ({\it middle panels}), and 60 $\mu$m infrared 
emission distributions ({\it right panels}) in the area the molecular shell \g. In the bottom panels the images  with point sources subtracted are shown. As in previous figures, the ellipse 
shows the outline of \g. }
\label{cont+60}
\end{figure*}

In order to study the diffuse continuum radio and infrared emission towards 
\g, we analyzed the 1420 and 408 MHz radio and \sixty\ infrared
images.
These images are shown in the upper panels of Figure \ref{cont+60},
 which shows a nearly $3^{\circ} \times 3^{\circ}$ field in the area of \gone.

The bright extended source partly visible in the top left corner of the radio images
is G\,127.1+0.5, a well known supernova remnant \citep{gre09}.  The diffuse emission 
observed at $l = 125^{\circ}$,  $b = 0^{\circ}$ is thermal in origin, as we show below.
The bright source
visible on all three images at $l = 126\fdg7$,  $b = -0\fdg8$ is
Sh2-187. It is immediately obvious that no significant continuum emission is
associated with \gone, although emission is indeed present but at a very low
level.  Both radio images are dominated by a very large number of 
point sources.   The radio image at 408 MHz has a very mottled appearance,
which suggests that the confusion limit is probably reached at this
frequency.

\begin{figure}
\includegraphics[width=8cm]{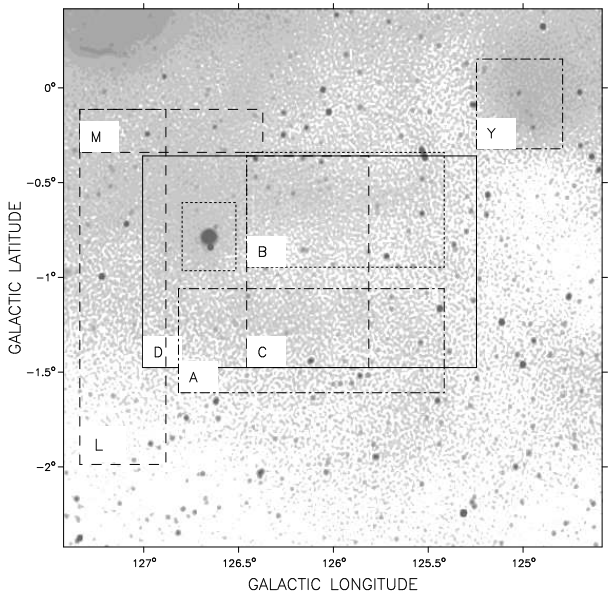}
\caption{CGPS continuum image at 1420 MHz showing the boxes used to calculate
the spectral indices using the TT-plot method. Labels are positioned at the
bottom left corner of each box.}
\label{figboxes}
\end{figure}

We first attempt to estimate the radio spectral index of the region. 
On both 1420 and 408 MHz images, point sources were first removed
using the program {\tt fluxfit} from the DRAO export software.  The resulting
images are shown in the bottom panels of Fig. \ref{cont+60}.  Some
artifacts of the removal process are present at 408 MHz (this is partly due
to the mottled nature of the initial image -- middle top panel in Fig.~\ref{cont+60}).
The next step consists in smoothing the two images at the same resolution.  Since the
region of diffuse emission covers a very large area, we chose to smooth to a common
resolution of 6\am.  After regridding to a coarser grid to avoid oversampling, we
carried out a TT-plot analysis of the region over a number of different sub-areas
which are shown in Figure \ref{figboxes}.  The dotted box around Sh2-187 outlines
the area which was excluded from the analysis. Boxes A to D enclose the emission
spatially coincident with \gone, whereas boxes L and M are control regions outside
this emission.  Box Y covers the extended diffuse emission seen in the top right
corner of the images.

\begin{figure*}
\includegraphics[width=13cm]{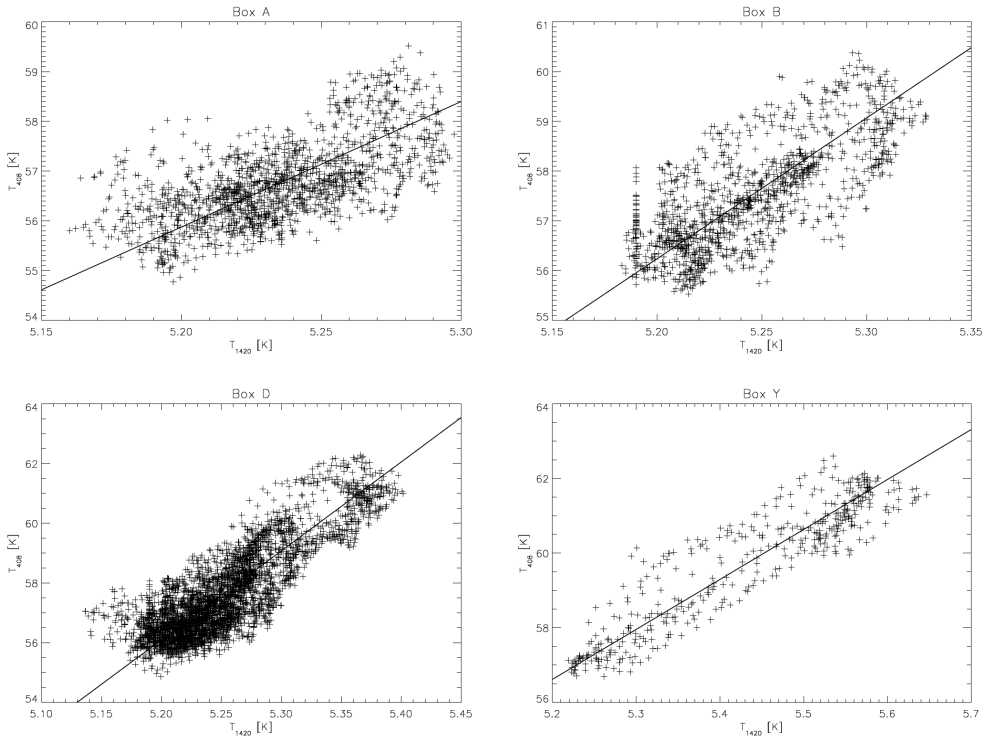}
\caption{TT-plots between 1420 and 408 MHz obtained for selected areas. The corresponding box is indicated at the top of each plot.}
\label{figttplots}
\end{figure*}

The temperature spectral indices $\beta$
[$\beta$ is defined by $T_B \propto \nu^{-\beta}$; the usual intensity spectral index
$\alpha$, where $S_{\nu} \propto \nu^{-\alpha}$, follows from $\alpha = \beta -2$] 
are summarized in Table \ref{tabspecind}  and
representative TT-plots shown in Figure \ref{figttplots}.
Theoretical values would be $\beta = 2.1$ for an optically thin \hii\, region, and around 2.5 for a SNR.

Although the faintness of the emission results in somewhat scattered plots
and hence relatively large uncertainties, all values for the region coincident
with \gone\ are consistent with non-thermal emission.  Note however that the two
control regions to the left and top left of \gone\ also show a non-thermal spectrum.
The spectral index of region Y is definitely that of an \hii\, region.  As an additional
test, we did a TT-plot of the southernmost part of the SNR G127.1+0.5 \citep{gre09}
visible near the top left of our images. We obtain a spectral index $\alpha = 0.40 \pm 0.07$.
This compares well with the value of $0.46 \pm 0.01$ obtained by \cite{lea06} for the
total remnant (our relatively larger error arises from the fact that we only considered
the southernmost fainter part of the SNR and have considerably fewer points).

Further insight into the nature of the diffuse emission can be obtained by examining the
correlation between IR emission and the radio continuum.
Indeed, as has been shown by a number of authors \citep[e.g.][]{bro89, fur87},
a strong correlation exists for \hii\, regions between the radio brightness 
temperature and the $60\,\mu$m intensity.  We attempted to reproduce such
correlation by following the procedure outlined by \cite{cic01} and
\cite{cap02}.  However the correlation coefficients are highly uncertain, reflecting
the fact that the diffuse emission is very weak, and the errors on
the slopes of the regression analysis are so large that is is not possible to
distinguish between a thermal or non-thermal nature of the radio emission.

 In summary, bearing in mind that aging SNRs see their surface brightness decrease with time as the expanding
structure gets spread over ever increasing volumes and the relativistic
electrons leak out and/or keep losing their energy,
the signature of a very old SNR is therefore likely to
be the presence of a lower density cavity having a lower
surface brightness, which is what we are observing here. The non-thermal
spectral index (or equivalently the absence of any significant thermal
component) is consistent with this interpretation.

\begin{table}
\caption{Radio spectral indices of selected areas}\label{tabspecind}
\vskip 0.25truecm
\centering
\begin{tabular}{ccccccc}
\hline\hline
Area  & $\beta_1$  & $\Delta \beta_1$ & $\beta_2$  & $\Delta \beta_2$\\
\hline
A  &  2.59  & 0.36  & 2.63  &  0.67  \\
B  &  2.68  & 0.25  & 2.70  &  0.32  \\
C  &  2.75  & 0.40  & 2.69  &  0.69  \\
D  &  2.72  & 0.15  & 2.77  &  0.16  \\
L  &  2.42  & 0.07  & 2.43  &  0.08  \\
M  &  3.20  & 0.18  & 3.21  &  0.21  \\
Y  &  2.08  & 0.07  & 2.06  &  0.07  \\
\hline
\end{tabular}
\begin{list}{}{}
  \item $\beta_1$ and  $\beta_2$ are obtained from the bisectrix method
   \citep{iso90,fei92} and the specindex
    routine of the DRAO export software, respectively, and $T_B \propto
    \nu^{-\beta}$.
\end{list}
\end{table}

\section{Discussion}

\subsection{Formation scenarios for \g}

In this Section, based on the estimated parameters of the molecular structure, we attempt to elucidate its possible origin. 
Knowing the importance that massive stars have in shaping the interstellar medium, 
we analyze two possibilities for the origin of the shell; stellar winds from OB stars and/or SN explosions.
To analyze the first possibility, we look for massive stars located in the area of \g.

\subsubsection{Stellar content inside the shell}\label{st}

We searched the available literature for identifying the population of massive star progenitors of
this cavity. The Simbad database resulted in 11 stars with spectral types B (see Table \ref{stars}). 
We also queried  the Early-Type Emission-Line Stars Catalogue \citep{wac70}, the 
Catalogue of Be stars \citep{jas82}, and the 
H-alpha Stars in the Northern Milky Way Catalogue \citep{koh97} for early-type, Wolf-Rayet, and emission stars. All of the found targets were
already included in the Simbad results.

\begin{table*}
\caption{Main parameters of massive star candidates to be responsible of \g.}\label{stars}
\vskip 0.25truecm
\centering
\begin{tabular}{lccccccccc}
\hline\hline
MAIN\_ID	&ALS	&Galactic Long. & Galactic Lat. &B	&V	&SP\_TYPE	&refs.	&new SP\_TYPE	&Distance\\
                &  & (deg.) & (deg.)    &(mag)  &(mag)  &               &       &               &(kpc)\\
\hline
BD+60   203&	&126.1946 &	--1.1261 		&9.64	&9.51	&B9 V	& $a$ 	&	&0.5\\
BD+60   200&	&126.0411 &	--0.9714 		&9.82	&9.7	&B5	& 	&	& 0.9\\
TYC 4030-98-1	&	&126.1901 &--0.8636 &12.16	&11.45	&B3	& $b$	&	& 1.0\\
BSD   8-1725	&& 125.74 &	--1.08 	&12.25	&11.69	&B7	& 	&	& 1.0\\
TYC 4030-352-1	&	& 125.6700 &--1.0652 	&11.76	&11.41	&B5n	&$e$	&	& 1.4\\
TYC 4034-902-1	&&126.0620 &	--0.7879	&11.3	&11.2	&B8	& 	&	& 1.6\\
HD   7720	&6520	&126.1082 &--0.8121 &9.37	&8.86	&B5 II e	& $c, d$ &&$>$1.6\\
TYC 4034-344-1	&6511	&125.8732 &--0.7827 &10.65	&10.31	&B5	& 	&B1.5 Ve	& $>$1.8\\
EM* MWC  422	&6498	& 125.6813 & --0.8134 &11.05	&10.26	&B0	&$b$	&B2 II e	& $>$2.4\\
EM* CDS  141	&6549	&126.5008 &--0.7153 &11.98	&11.12	&B...	& 	&B0.5 IIIe	&$>$ 3.3\\
TYC 4030-1547-1	&6555	& 126.6284 &--1.0988&11.4	&10.9	&Bpe	& $d$	&B1 III e	& $>$ 3.7\\
\hline
\end{tabular}
\begin{list}{}{}
 \item References: {$a$) \citet{feh61}, $b$) \citet{pop44}, $c$) \citet{hil56}, $d$) \citet{jas82}, $e$) \citet{sey41}}. 
\end{list}
\end{table*}

As can be noted in Table \ref{stars}, some stars have an uncertain spectral type:\\
\begin{itemize}

\item ALS6511 is labeled as B5 D in Simbad, but no reliable cite is provided nor independently found by us.
\item ALS6555 is a known Be-type star discovered by \citet{mer50}, and included in the Catalogue of
Be stars \citep{jas82} but no precise sub-type is provided.
\item ALS6498 is labeled as B0 D in Simbad, but the unique classification found is a rough estimation,
B(4)ne, by \citet{pop44}.
\item ALS6549 is labeled just as B D in Simbad, but no reliable cite is provided nor independently found
by us.
\item BSD 8-1725 seems to be misidentified thus was not considered.
\item BD +60 200 is labeled as B5 in Simbad. No cite was found.
\item TYC 4034-902-1 is labeled as B8 in Simbad, but no reliable cite is provided nor independently
found by us.
\end{itemize}

Thus, we observed the spectrum of the most interesting ones and classified them following the
criteria for OB stars \citep{wal90}. The resulted new spectral types are
shown in Column 9 of Table \ref{stars}.

We determined the spectroscopic distances for each of the stars
in Table \ref{stars}, and these are given in Column 10 of the Table. 
The visual absorption $A_V$ was determined as $A_V$ = R$_V 
\times$ E$_{B-V}$. The value of $R_V$ was adopted  to be consistent with the distance to Sh2-187 determined
 by \citet{rus07}, who used R$_V$ = 3.2. The intrinsic colors of the B-type stars were extracted from the
\citet{weg94} calibrations. We have also discarded the fact that 
Be stars are intrinsically redder than B-normal stars, thus these distances should be considered as minimum ones.

It is noteworthy that six of the 11 stars share a common distance of  1.4 $\pm$ 0.3 kpc
(error is represented by the standard deviation of the mean) which is similar to the value
determined from the CO emission analysis (1.44 kpc; see Sect. \ref{co}).

Naturally, we consider these six B stars as part of the stellar population inside the shell.
 Hence, we estimate whether the mechanical energy input provided by these stars may be sufficient
enough to create \g. 
The earliest spectral type among these six stars is B1.5Ve, so, for numerical purposes, we shall consider the six stars as B1V, and obtain an upper limit for the injected  wind energy.
Considering for a B1V star a wind speed of 2500 \kms, and a mass loss rate of log($\dot{M}$) = $-8.2$ \citep{lei98}, 
the estimated wind luminosity is about 1.3 $\times$ 10$^{34}$ erg\,s$^{-1}$.
Then, the total wind energy, $E_w$, injected by the six stars  since the creation of the shell  ($\sim 1.7 \times 10^6$ yr) is $E_w < 4.2 \times 10^{48}$ erg,
significantly lower than the kinetic energy in the shell determined in Sect. \ref{co}, $E_{\rm kin} =$ (1.3 $\pm$ 0.8) $\times 10^{49}$ erg).  Moreover, bearing in mind that theoretical models predict that only 20\% of the wind energy is converted into mechanical energy of the shell \citep{wea77}, and that the observational analysis of several  \hi\, shells shows that the energy conversion efficiency is in fact  lower, roughly about 2–-5\% \citep{cap03} it is clear that the origin of \g\, cannot be due to the action of only the winds of the  B stars.

 It is worth mentioning that the conversion factor  $X=N({\rm H_2})/W_{{\rm CO}}$  used in Section \ref{co} to estimate the mass of H$_2$, and then the kinetic energy of the shell, introduces unknown uncertainties which are hard to quantify since we are using a statistical relation for a single object or line of sight.
However, the numbers we obtain for the energy injected by the stars cannot explain the origin of the shell even taking into account a 50\% error in the adopted value of $X$.

\subsubsection{A supernova remnant?}

As the energy injected by the stars observed inside \g\, is not enough to create such a structure on its own, we now analyze if the shell is likely the result of a supernova explosion (SN$_e$).

Based on the dynamical age derived for \g, 
 (0.7-1.7) $\times 10^6$ yr  (see Sect. \ref{hi}), this structure certainly has outlived any observational evidence (e.g. a {\rm SNR}) that may have had its
 origin in a SN$_e$ involved in the genesis of \g.
In this sense, the faint radio continuum emission observed in the area of \g\,, which seems to have a non-thermal origin (see Section \ref{rc}), could be interpreted as the fingerprint of a putative SNR.

The required SN explosion energy to produce \g\, can be estimated as $E_{\rm}= 6.8 \times 10^{43}\, n_0^{1.16}\, R^{3.16}\, v_{\rm exp}^{1.35} \, \psi^{0.161}$ erg, where $n_0$ is the ambient H density in cm$^{-3}$, $R$ is the radius of the shell in pc, $v_{\rm exp}$ is the expansion velocity in \kms, and $\psi$ is the metalicity in units of the solar value \citep{cio88}.
As a rough estimate, 
the original ambient density can be estimated by distributing the total mass (ionized, neutral atomic and molecular) 
related to the structure over its volume. Adopting a spherical geometry with $R = 13$ pc we obtain  $n_o = 285 \pm 130$ 
cm$^{-3}$. Assuming a solar metalicity, we estimate 
$E_{\rm}= (1.2  \pm 0.8) \times 10^{51}$ erg, which,  agrees with the canonical value of the energy injected by a single SN explosion ($ 10^{51}$ erg). 
 The  difference between this value and the kinetic energy of the shell estimated in Section \ref{co}, $E_{\rm kin} =$ (1.3 $\pm$ 0.8) $\times 10^{49}$ erg, is consistent with the shell being an old SNR where most of the input energy has already been dissipated through radiative losses  and only a small percentage of it  remains as kinetic energy of the expanding gas \citep{che74}.

Looking for evidence that such an explosion might have taken place
 in the past, we searched for pulsars seen in projection against \g\, within a circle of $5^{\circ}$ radius centered at ({\it l, b}) = (126\fdg68, --0\fdg82).
Those pulsars that might be associated with the SN$_e$ that may have  contributed to the formation of \g\, must have both an age comparable to, or lower than, the dynamical age of \g, and a pulsar distance in agreement, within a 2$\sigma$
uncertainty, with the distance of \g. Searching the {\rm ATNF} pulsar
database\footnote{http://www.atnf.csiro.au/research/pulsar/psrcat/} \citep{man05}, imposing the restrictions indicated above, it turns out that 
two pulsars, namely B0105+65 and B0154+61, fulfilled this set of restrictions. 
The pulsars are listed in Table \ref{table:pulsar}.

\begin{table}
\centering
\begin{minipage}{140mm}
\caption{ Pulsar main parameters.}
\begin{tabular}{lllll}
\hline\hline  
 Name & Galactic Long. & Galactic Lat. & Distance & Age\\
      &  (deg.) & (deg.) & (kpc) & (10$^6$ yr) \\    
\hline 
  B0105+65 & 124\fdg646 & 3\fdg327 & 1.42 & 1.56\\
  B0154+61 & 130\fdg585 & 0\fdg326 & 1.71 & 0.197 \\
  \hline
\label{table:pulsar} 
\end{tabular}
\end{minipage}
\end{table}

Based on the quoted age, only B0105+65 could have been related to the SN$_e$ that might have played a role in the formation of \g. Were this so, a minimum pulsar space velocity of $\sim$73 kms$^{-1}$ could be derived under the assumption that the pulsar is mostly moving in the plane of the sky. Assuming a modest
 velocity component along the line of sight between 25 and 75  \kms, the space velocity would be increased to $\sim$ 77 and $\sim$ 105  \kms, respectively. Both figures are within the low velocity tail of the three-dimensional velocity distribution of pulsars quoted by \citet{hob05}.

In the case of B0154+61, its relative youth compared to the dynamical age of \g\, implies that the SN$_e$ from which B0154+61 was born played
no role in the genesis of \g, though it could have played a role in 
maintaining the shell expansion. In this case the minimum space
velocity needed for B0154+61 to reach its present position, under the assumption it was born at the very center of \g, is $\sim$ 508 \kms. Though
high, this spatial velocity falls on the high velocity tail of the distribution 
found by \citet{hob05}. In the very likely event that B0154+61 has a
 non-zero radial velocity, a maximum approaching (or receding) radial velocity of $\sim$ 740 \kms\, is still compatible with the very high spatial velocity tail of the distribution published by \citet{hob05}.  \\

In summary, a supernova explosion seems to be the origin of \g. It is important to mention, however, that
given the huge influence that massive star winds have in the ISM, 
the possibility that the progenitor of the SNe
played an important role in shaping the surrounding molecular gas cannot be discarded.

\subsection{Triggered star formation in the shell?}

As a swept-up molecular shell may harbour a new generation of stars \citep{elm98}, we examine the young stellar object candidates (cYSO) distribution in the molecular structure, making use of the available  
 point-source infrared catalogs.

It is well known that YSOs show excess emission at infra-red (IR) wavelengths due to thermal emission from their circumstellar material. Thus, YSOs can be identified by looking for objects showing IR excess emission. Furthermore, IR data can also be used to classify YSOs at different evolutionary stages, which may help to probe the presence of recent star formation activity in a given region. 

\subsubsection{YSOs classification}

To look for primary tracers of stellar formation activity we used the MSX Infrared Point Source Catalogue \citep{ega03}, the IRAS Point Source Catalogue \citep{bei88}, 
and the WISE  All-Sky Source Catalogue
\citep{wri10} .
Within an area of size  $\Delta l \times \Delta b$ = 1\fdg8 $\times$ 1\fdg3, centered at ($l, b$) = (126\fdg1, --0\fdg85), a total of 39 IRAS, 112 MSX, and 2617  WISE  sources were found.

To identify the cYSO among the IRAS sources, we applied the \citet{jun92} color criteria: $F_{100} \geq$ 20 Jy, 1.2 $\leq F_{100}/F_{60} \leq$ 6.0, $F_{60}/F_{25} \geq$ 1, and $Q_{60} + Q_{100} \geq$ 4, where $F_{\lambda}$ and $̣Q_{\lambda}$ are the flux density and the quality of the IRAS flux in each of the observed bands, respectively. Only two IRAS sources out of the 39 catalogued in the area meet the criteria and could be identified as protostellar candidates. Their properties are listed in Table \ref{ysos}.

As for the MSX sources, only 8 out of the 112 have  acceptable flux quality  (q $\geq$ 2) in the four  bands. They were classified according to the \citet{lum02} criteria using the 
 flux densities (F$_{\lambda}$) in each of the four bands (A (8.3 $\mu$m), C (12.1 $\mu$m), D (14.7 $\mu$m), E (21.3 $\mu$m)). According to these criteria, massive young stellar objects (MYSO) candidates should have F$_{21}$/ F$_{8} \geq 2$ and F$_{14}$/ F$_{12} \geq 1$, while compact \hii\, regions (C\hii) should have  F$_{21}$/ F$_{8} \geq 2$ and F$_{14}$/ F$_{12} \leq 1$. We found 3 MYSO  and 2 C\hii\, candidates. Their properties are listed in Table \ref{ysos}.

\begin{table*}
\small
\caption{ IRAS and MSX YSO candidates projected onto the area of \g. \label{ysos}}
\begin{tabular}{c c l l l l l l}
\hline
\multicolumn{8}{c}{{\bf IRAS sources}} \\
\hline
 $\#$ &  Designation &  {(\it l, b})&  F$_{12}$[Jy] (Q$_{12}$) &  F$_{25}$[Jy] (Q$_{25}$) &  F$_{60}$[Jy] (Q$_{60}$) &  F$_{100}$[Jy] (Q$_{100}$) & Notes \\
\hline
1& 01195+6136 & 126\fdg62, --0\fdg77& 7.18 $\pm$ 0.78 (3)& 7.99 $\pm$ 1.04 (3)& 700.8 $\pm$ 56.1 (2)& 1953.0 $\pm$ 214.8 (3)&  \\
2 & 01202+6133 & 126\fdg71,  --0\fdg82 & 10.44 $\pm$ 0.83 (3)& 182.3 $\pm$ 7.3 (3)& 881.5 $\pm$ 52.9 (3) & 1710.0 (1: upper limit)& \\
\hline
\multicolumn{8}{c}{{\bf MSX sources}} \\
\hline
 $\#$ &  Designation &  {(\it l, b})&  F$_{8}$[Jy] (Q$_{8}$) &  F$_{12}$[Jy] (Q$_{12}$) &  F$_{14}$[Jy] (Q$_{14}$) &  F$_{21}$[Jy] (Q$_{21}$) & Notes \\
\hline
3& G126.6517-00.7799 & 126.652,  --0.78 & 0.52 (4)& 1.43  (2) & 1.17 (2) &  5.29 (4) & C\hii\\
4& G126.6777-00.8115 & 126.678, --0.81 & 0.46 (4)&  1.47 (2) &  1.11 (2) &   2.27 (2) &C\hii\\
5& G126.4274-01.2348 & 126.427, --1.235 & 1.2 (4) & 4.12 (4) & 6.32 (4) & 10.16 (4) & MYSO\\
6& G126.6541-00.7828 & 126.654, --0.782 &0.796 (4)& 1.69  (3) & 2.03  (4)&  5.078 (4) &MYSO\\
7& G126.7144-00.8220 & 126.714, --0.822 & 8.21 (4) & 12.64 (4) & 22.96  (4)& 104.52  (4) & MYSO\\
\hline
\end{tabular}
\end{table*}

\begin{table*}
\scriptsize
\caption{ WISE + 2MASS YSO candidates projected onto the area of \g. \label{wise}}
\begin{tabular}{c c c c c c c c c c l}
\hline
\hline
$\#$ &  Designation &  {(\it l, b})&  W1 [mag]  &    W2 [mag]   &  W3 [mag]  &  W4 [mag] & $J$[mag] & $H$[mag]& $K_s$[mag] & Notes \\
\hline
8 & J012254.26+621510.9 &126\fdg58, --0\fdg395 & 13.651 $\pm$ 0.071 & 10.931 $\pm$ 0.033 &  8.189 $\pm$ 0.033 & 3.694 $\pm$ 0.029 &   NA  &   NA &    NA& Class I\\ 
9 & J012142.12+622535.3 & 126\fdg426, --0\fdg24 & 10.101 $\pm$ 0.023 &  7.992 $\pm$ 0.020 & 5.554 $\pm$ 0.014 & 3.045 $\pm$ 0.014&  16.474(U)& 15.113 (B)&12.832(A) & Class I; CO at $ -59$\kms.\\
10 & J012123.30+622857.4 & 126.38, --0\fdg188 & 12.139 $\pm$ 0.023 & 11.012 $\pm$ 0.021 & 8.682 $\pm$ 0.031 &   6.554 $\pm$ 0.073& 16.461(B)& 15.018 (A)&14.025(A) & Class I; CO at $-59$\kms. \\ 
11 &J012138.90+622734.3 & 126\fdg416, --0\fdg207 & 13.558 $\pm$ 0.028 & 12.245 $\pm$ 0.026 & 9.665 $\pm$ 0.052 & 6.455 $\pm$ 0.058 &16.394(B) &15.165(A) &14.399(A)& Class I; CO at  $-59$\kms.\\
12 & J012249.24+622145.5 & 126\fdg562, --0\fdg287 & 12.310 $\pm$ 0.024 &10.257 $\pm$ 0.020& 7.926 $\pm$  0.029& 5.134 $\pm$ 0.032& 18.063(U)&16.785(U) &14.477(A)&Class I \\
13 &J011441.19+621103.2 &125\fdg64, --0\fdg565& 11.251 $\pm$ 0.023 &10.169 $\pm$ 0.020 &8.014 $\pm$ 0.023& 5.909 $\pm$ 0.040& 16.998(D)& 15.033(A)& 13.525(A)& Class I\\
14 & J012211.23+615352.1 &126\fdg54, --0\fdg75& 15.716 $\pm$ 0.113& 14.375 $\pm$ 0.098 &8.944 $\pm$ 0.035& 6.469 $\pm$ 0.085&  NA  &   NA &    NA &Class I\\
15 &J011612.44+613537.8 &125\fdg873, --1\fdg136 &13.451 $\pm$ 0.028& 12.356 $\pm$ 0.025 &9.977 $\pm$ 0.051 &7.539 $\pm$ 0.117& 18.527(U) &15.793(U)& 15.301(C) &Class I, CO at $ -48$\kms.\\
16 &J011918.34+621929.7& 126\fdg161, --0\fdg371& 10.602  $\pm$ 0.021& 10.187 $\pm$  0.020 & 5.701 $\pm$  0.015& 2.668  $\pm$ 0.015&13.668(U) &12.448(U) &12.093(A)& Class II - Galaxy$^a$\\
17 & J012256.12+621631.0 & 126\fdg586, --0\fdg372  &11.543  $\pm$ 0.024  &10.972 $\pm$  0.022 &  9.746 $\pm$ 0.058 & 7.975  $\pm$ 0.201& 14.302(A)& 13.003(A) &12.330(A)&Class II\\
18  & J012315.69+620522.5 & 126\fdg646, --0\fdg552 & 12.250 $\pm$ 0.024 & 11.459 $\pm$  0.022  & 9.593 $\pm$ 0.039 & 7.325 $\pm$  0.092& 15.877(A)& 14.312(A) & 13.239(A) & Class II\\
19 &  J012216.46+621448.4 & 126\fdg513, --0\fdg410 & 11.297  $\pm$ 0.023 & 10.896 $\pm$  0.022 &  9.285 $\pm$   0.032 & 7.366 $\pm$  0.093& 13.348 (A)& 12.333(A) &11.885(A) &Class II\\
20 & J012033.50+623350.7  &126\fdg279, --0\fdg118 & 13.182  $\pm$ 0.025 & 12.441 $\pm$  0.026 & 10.160 $\pm$  0.075 & 7.446  $\pm$ 0.136 & 15.841(U) &15.598(C) &14.486(A)&Class II\\
21 &  J012247.68+621816.1 & 126\fdg566, --0\fdg345 &  9.429 $\pm$  0.024 &  8.763 $\pm$ 0.020 &  6.891 $\pm$ 0.019 & 4.876  $\pm$ 0.030 &12.203(A)& 10.988(A) &10.298(A) &Class II\\
22 &  J012240.35+622325.1  &126\fdg542, --0\fdg262 & 12.125 $\pm$ 0.025 & 11.752 $\pm$  0.024 & 10.097 $\pm$  0.056 & 7.687 $\pm$  0.135 &13.958(A) &12.962(A)& 12.552(A)&Class II\\
23 &  J012250.42+622227.0  &126\fdg563, --0\fdg275 & 10.932 $\pm$ 0.023 & 10.047  $\pm$ 0.020 &  7.798 $\pm$  0.022 & 5.654 $\pm$  0.034 &15.382(A) &13.756(A) &12.612(A)&Class II\\
24 & J011410.05+620729.9  &125\fdg585, --0\fdg629  & 9.695 $\pm$ 0.023 &  8.840  $\pm$ 0.020  & 6.535 $\pm$  0.016 & 4.224 $\pm$  0.019 & 12.964(A) & 11.818(A)& 10.973(A)& Class II\\
25 & J012248.12+613422.7  &126\fdg656, --1\fdg071 & 13.186 $\pm$ 0.027 & 12.883  $\pm$ 0.032 & 10.818 $\pm$  0.104 & 7.963  $\pm$ 0.180 &15.084(A) &14.217(A)& 14.174(A)&Class II\\
26 & J012231.37+613407.4  &126\fdg623, --1\fdg079 & 13.828 $\pm$  0.029 & 13.325 $\pm$  0.035 & 10.622 $\pm$  0.086 & 7.733  $\pm$ 0.146 &15.488(A) &14.682(A)& 14.207(A)&Class II\\
27 & J012307.47+613322.7 & 126\fdg696, --1\fdg083 & 12.381 $\pm$ 0.027 & 11.918 $\pm$  0.024 & 10.301 $\pm$  0.072 & 8.204  $\pm$ 0.220 &14.099(A) &13.178(A)& 12.835(A)&Class II\\
28 & J011530.65+613730.7 & 125\fdg788, --1\fdg112 & 13.276 $\pm$  0.027 & 12.663 $\pm$  0.028 & 10.128 $\pm$  0.053 & 7.323 $\pm$  0.105 &16.925(C) &15.693(B)&14.430(A)&Class II; CO at $-48$\kms.\\
29 & J011553.88+613336.9 & 125\fdg840, --1\fdg173 & 12.812  $\pm$ 0.025 & 12.061 $\pm$  0.026 &  9.174 $\pm$  0.034 & 6.648 $\pm$  0.073 & 14.621(A) &14.063(A)& 13.773(A) &Class II; CO at $-48$\kms.\\
30 & J011602.88+613513.7 & 125\fdg855, --1\fdg144 & 13.471 $\pm$  0.027 & 12.547 $\pm$  0.027 &  9.852 $\pm$  0.044 & 7.387 $\pm$  0.098 & 16.531(B) &15.398(B) &14.509(A)&Class II; CO at $ -48$\kms.\\
31 & J012052.04+611953.0 & 126\fdg455, --1\fdg338 & 11.723 $\pm$  0.023 & 11.153 $\pm$  0.022 &  8.555 $\pm$  0.022 & 6.188  $\pm$ 0.049 &13.813(A) & 12.967(A)& 12.508(A)& Class II\\
32 & J012039.92+610830.6 & 126\fdg452, --1\fdg529 & 12.129 $\pm$  0.024 & 11.475 $\pm$  0.022 &  9.486 $\pm$  0.033 & 7.315 $\pm$  0.086 &15.369(A) &14.116(A)& 13.271(A)&Class II\\
\hline
\end{tabular}
\begin{list}{}{}
 \item {$^a$} This source coincides with IRAS\,01159+6203 and  with 2MASX\,J01191829+6219297, and is part of the 2MASS-selected flat galaxy catalog \citep{mit04}. 
\end{list}
\end{table*}

\begin{figure}
\includegraphics[width=8cm]{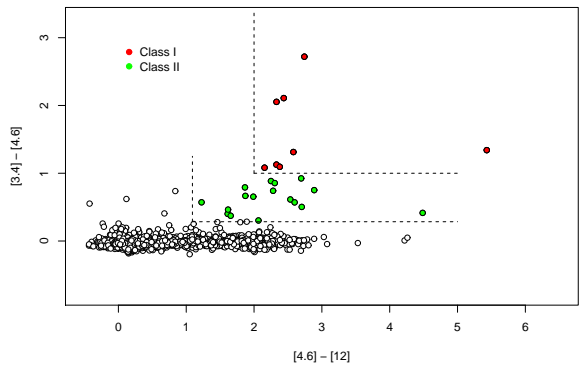}
\caption{ WISE color-color diagram showing the areas where the Class I and Class II YSO candidates are located according to the \citet{koe12} criteria. }
\label{cc}
\end{figure}

To classify the WISE sources we adopted the  classification scheme described in \citet{koe12}.  The WISE bands are often referred to as W1 (3.4 $\mu$m), W2 (4.6 $\mu$m), W3 (12 $\mu$m), and W4 (22 $\mu$m). We selected all sources having in all four bands a photometric uncertainty $\le$ 0.2 mag  and S/N $\ge$ 7.  
As mentioned by \citet{koe12}, various non-stellar  sources, such as PAH-emitting galaxies, broad-line active galactic nuclei (AGNs), unresolved knots of shock emission, and PAH-emission features, must
be eliminated, using different color criteria, from the listed sources before attempting to identify the cYSOs.
The sources dropped from the listing get the global name of contaminants.  A total of  822 contaminants were found. The remaining sample of 1795 sources were used to identify Class I (sources where the IR emission arises mainly from a dense infalling envelope) and Class II (sources where the IR emission is dominated by the presence of an optically thick circumstellar disk) 
YSO candidates to be associated with \g\, using the criteria given by \citet{koe12}.
In total, we have identified 8 sources with protostellar colors (Class I) and 17 with colors consistent with a pre-main-sequence star having  a circumstellar disk (Class II).
In Fig. \ref{cc} we show the (W2 -- W3) vs (W1 -- W2) color-color diagram for all the uncontaminated sources, where the Class I and Class II sources are shown in red and green, respectively.
For the cYSOs, we check whether their W4 magnitudes are consistent with the Class I/II classification, i.e, that the sources previously classified as Class I have rising SEDs at 22 $\mu$m and that the Class II sources do not present excessively blue colors.  
None of the 25 cYSOs has to be rejected.
On the other hand, protostellar objects with intermediate/high masses can be identified among the Class I sources by additionally requiring that W3 $<$ 5 \citep{hig13}. None of the 8 Class I sources in \g\, satisfies this criterion, suggesting that they are low mass protostars.

For each cYSO, JHK photometry, when available, was obtained from the 2MASS catalogue. Table \ref{wise} summarizes the cYSO sample photometry. The corresponding photometric quality flag of each of the  2MASS magnitudes are given between parenthesis.

\subsubsection{Spatial distribution and main properties of cYSOs}

\begin{figure}
\includegraphics[width=9cm]{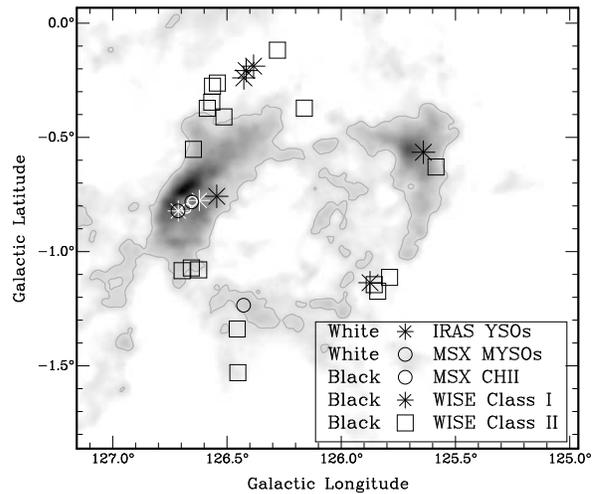}
\caption{Infrared point sources classified as cYSOs overlaid on to the CO distribution averaged between the velocity range from --6.4 to --19.6 \kms. }
\label{ysos-dist}
\end{figure}

In Fig. \ref{ysos-dist} we show the spatial distribution of all the identified cYSOs overlaid on the averaged CO image. Different symbols and colors indicate different classes of sources. 
As can be noticed, both IRAS sources and all but one MSX sources are located in the  area of Sh2-187.
As already pointed out by \citet{kun08}, this region presents several signs of recent star formation, such as an outflow source (S187 IRS) \citep{bal83}, H$_2$O maser emission \citep{hen86}, and  an optically visible young stellar object (S187H$\alpha$) \citep{zav94}. 
In particular, IRAS source \#\,2, which coincides with MSX source \#\,7,  is  located 3\arcmin\,  southeast  of Sh2-187 
and was found to be surrounded by an infrared nebula referred to as S187\,IR by \citet{hod94}. \\

The MYSO candidate \#\,5 is the unique MSX source seen projected onto \g\, but outside the area of Sh2-187. This infrared source coincides with IRAS\,01174+6110, the 2MASS source J01204420+6126158, and the WISE source J012044.25+612615.7.
It is important to mention that CO emission in this direction is only observed at the radial velocity interval between --9 and --15 \kms, coincident with the radial velocity of \g, suggesting an association between the MSX source and the molecular structure. This was already pointed out by \citet{ker03}, who associated this source with molecular gas at --12.12 \kms.

The nature of this source is  controversial. \citet{koh01} lists   IRAS\,01174+6110  in the Catalogue of Galactic Planetary Nebulae, while other works suggest that the source is a compact \hii\, region \citep{gar97,kel05}. Later, based on optical spectroscopy, \citet{sua06} classified IRAS01174+6110 as a young source, and was classified as an YSO in the RMS survey \citep{urq08}.

\begin{figure*}
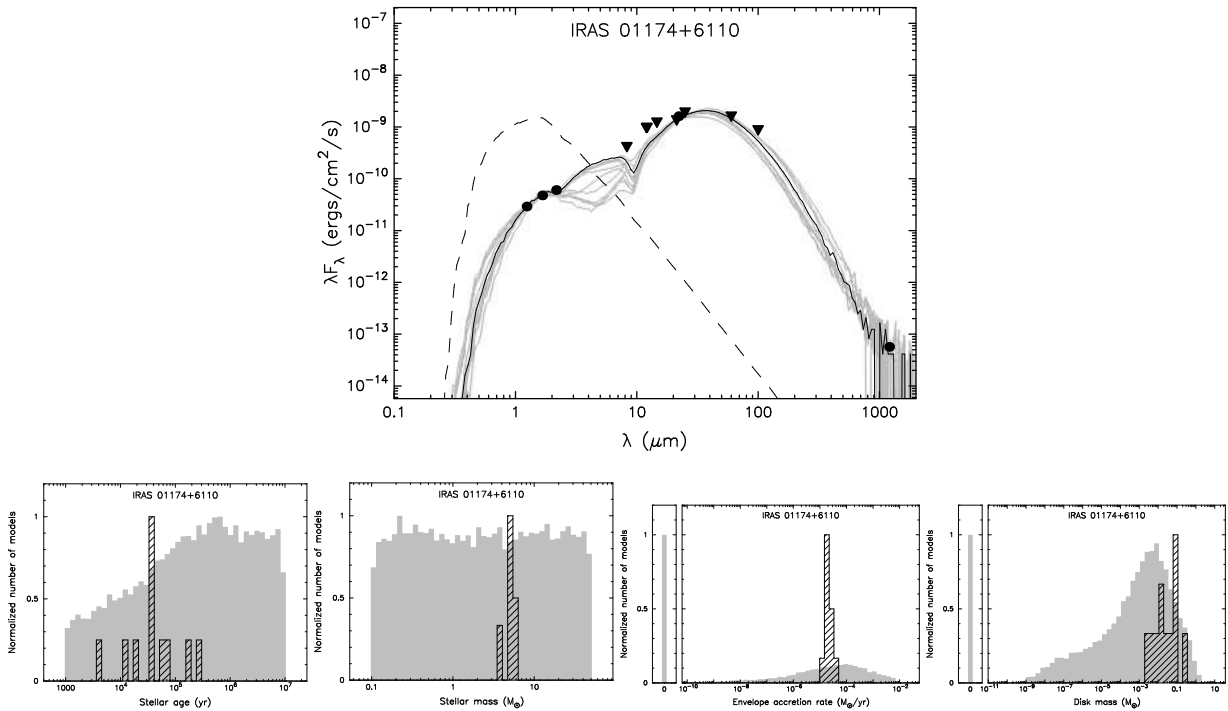

 \centering
  \subfloat{
   \label{sed}
    \includegraphics[width=8cm]{fig11-1.eps}}\\
\subfloat{
   \label{Stellar age}
    \includegraphics[width=4cm]{fig11-2.eps}}
  \subfloat{
   \label{mass}
    \includegraphics[width=4cm]{fig11-3.eps}}
  \subfloat{
   \label{acc}
    \includegraphics[width=4cm]{fig11-4.eps}}
\subfloat{
   \label{disc}
    \includegraphics[width=4cm]{fig11-5.eps}}
 \caption{IRAS\,01174+6110 SED. {\it Top panel}:The filled circles show the input fluxes and the triangles the ones that were set to upper limits  fluxes. The black line shows the best fit, and the gray lines show subsequent good fits. The dashed line shows the stellar photosphere corresponding to the central source of the best fitting model, as it would look in the absence of circumstellar dust (but including interstellar extinction). {\it Lower panels}: Stellar age, stellar mass, envelope accretion rate, and disk mass distributions.  The gray histogram shows the distribution of models in the model grid, and the hashed histogram shows the distribution of the selected models.}
 \label{sedhist}
\end{figure*}

To better characterize the source IRAS\,01174+6110, we constructed its SED  using the grid of models and fitting tools of \citet{rob06,rob07}. The SED fitting tool fits the data allowing the distance and external foreground extinction as free parameters. We gave a distance range of 1.0 - 1.4 kpc and an extinction range of 2 - 40 mag. We set photometric uncertainties to 10 \% for the 2MASS, WISE (only the W4 band was used for the fitting because the other three bands have either a non null contamination  or variability flag) and the 1.2mm IRAM data, and set as upper limit the MSX and IRAS inputs.
Figure \ref{sedhist}  shows the SED obtained for IRAS\,01174+6110, where the solid black line represents the best fit and the grey lines the subsequent well-fits, together with the distribution of some parameters obtained from the fitting, such as the stellar age, the stellar mass, the envelope accretion rate, and the disk mass. The distributions show the values  obtained for all  the fitted models satisfying $\chi^2 - \chi^2_{\rm best} < 2\, N$, where $\chi^2_{\rm best}$ is the goodness-of-fit parameter for the  best-fit model (we obtained $\chi^2_{\rm best}$ = 2.25) and  $N$ is the number of input observational data points ($N = 5$ in this case, which yields a number of well-fitted models of 11). 
Clearly the SED of IRAS\,01174+6110 corresponds to  a young stellar object rather than to an evolved source.
From the parameter distributions we can see that though the value of the stellar age is not well constrained, we can suggest that it is between 4 $\times 10^3$ and $3 \times 10^5$ years with a most probable value of $5 \times 10^4$ years. The other three parameters shown in the lower panels of Fig. \ref{sedhist} are better constrained and allow us to suggest that IRAS\,01174+6110 has an intermediate mass of 5.1 $\pm$ 0.7 M$_{\odot}$, an envelope accretion rate of (2.3 $\pm$ 0.8) $ \times 10^{-5}$ M$_{\odot}\,$yr$^{-1}$ and a mass disk between 0.002  and 0.2 M$_{\odot}$.

With regards to the WISE cYSOs, they are distributed in different parts of the shell and most of them are grouped (Fig. \ref{ysos-dist}). 
An inspection of the CO data cube shows that in direction to all the cYSOs CO is observed at the velocity of \g\,, suggesting a relation between the cYSO and \g.  A second CO velocity component is observed for two groups of sources. One corresponds to the  4 cYSOs located near ($l, b$) $\sim$ (125\fdg9, --1\fdg15), where CO is observed at --15 \kms\, but also at around --48 \kms.
At ($l, b$) $\sim$ (126\fdg4, --0\fdg2), where 3 cYSOS are located, CO is observed at --11 and --59 \kms.
The sources that present a second velocity component are indicated in the last column of Table \ref{wise}.

Based on the photometry data of Table \ref{wise}, 
we attempt to constrain the evolutionary status of the cYSOs and estimate some parameters using the SED fitting tool developed by \citet{rob07}.
As before, we fixed the minimum photometric error in any band to be 10\% to prevent a single data point
from dominating the resulting fits and to account for possible errors in absolute flux calibration \citep{rob07}.
The interstellar extinction was allowed to vary from 2 to 40 mag and the distance range was set to  1.0 -– 1.4 kpc. 
To estimate the parameters,  we carefully inspect every distribution (as in the case of IRAS\,01174+6110) considering all the models that satisfy the criteria $\chi^2 - \chi^{2}_{\rm best} < 2\, N$, where now $N = 7$. The results are listed in Table \ref{seds}. 
We found out that even when all the data points (for each of the 22 sources having WISE and 2MASS photometry)  can be fitted with several models with great goodness, just few parameters are well constrained and in some cases only  lower and/or upper limit values can be given.
Clearly, additional observational data points at longer wavelengths would help constrain the parameters more precisely.

\begin{table*}
\caption{ SED parameters for the WISE YSO candidates with 2MASS magnitudes. \label{seds}}
\begin{tabular}{c  c c c c c  c }
\hline
\hline
$\#$ &    $\chi^2_{\rm best}$ &  Nro. models &   Stellar Age  & Stellar Mass    & Envelope Accretion Rate   & Disk Mass   \\
&  & $\chi^{2} - \chi^{2}_{\rm best}$   $<$ 14 & yr & $M_{\odot}$ &  $M_{\odot}\, \rm yr^{-1}$ &  $M_{\odot}$  \\
\hline
9 &  0.14&10000 & $>10^4$ &3.9 $\pm$ 1.9  & $<10^{-4}$ & 	\\
10 & 0.13 & 10000 & $> 10^4$	&  $<$ 5.0& $<10^{-4}$	 & 	\\
11  &7.62 &23 &$10^4 - 4\times 10^5$	& $<$ 1.5& $< 10^{-5} $ &$10^{-5} - 0.02$\\
12& 14.51 &174&$>10^5$ &2.3 $\pm$ 0.9 & $< 5 \times 10^{-4}$	 & $5 \times 10^{-5} - 0.1$\\
13 &6.69 &16 & (7.7 $\pm$ 2.5) $\times 10^6$  & 2.2 $\pm$ 0.3 &0& $8 \times 10^{-6} - 0.007$ \\
15 & 0.06 &1831& $> 10^5$  &$<$ 3.4& $< 4 \times 10^{-5}$ & \\
17 & 1.56 &  1204 & $> 3 \times10^5$ & 1.4 $\pm$ 0.6 & $< 10^{-6}$ &  \\
18 & 1.65& 716 & $> 2 \times 10^4$ & $< 3.0$ & $< 4 \times 10^{-5}$ & $10^{-5}-0.1$\\
19 &0.22& 2941 & $> 10^5$ & 1.2 $\pm$ 0.6 & $<5 \times 10^{-5}$ & \\
20 & 0.54& 816 & $> 5 \times 10^4$ & $< 3.3 $ $< 4 \times 10^{-5}$ & $10^{-5}-0.1$\\
21&0.87 & 461 & $> 1.3 \times 10^5$& 3.1 $\pm$ 0.6 & $<10^{-5}$ & \\
22& 0.98& 705 & $>10^5$ & $<2.5$ & $< 5 \times 10^{-5}$ & $10^{-5}-0.1$ \\
23& 1.78& 79 & (5.6 $\pm 2.4$) $\times 10^6$ & 2.5 $\pm$ 0.6 & 0 & 	\\
24& 1.0 & 127 & (4.1 $\pm$ 2.2) $ \times 10^6$ & 2.6 $\pm$ 0.5& 0 & $10^{-5}-0.1$ \\
25 &12.98 & 40 & (1.8 $\pm 1.0$)$ \times 10^5$ & $0.15 \pm 0.04$ & $< 2\times 10^{-5}$& $3 \times 10^{-6}-0.007$\\
26 &1.35 & 87 & $> 4.4 \times 10^4$& $< 1.5$ & $< 2 \times 10^{-5}$ & $10^{-5}-0.006$ \\
27 & 0.97 &1340 & $>10^5 $ & $< 2.6 $& $< 5 \times 10^{-5}$& $10^{-6}-0.04$\\
28 & 8.13 &46 &	$>2 \times 10^4$ & $< 3.3$ & $ < 4 \times 10^{-5}$& $10^{-5}-0.008$\\
29 &16.94 &8 & ($1.3 \pm 0.9$)$\times 10^5$& $< 0.7$ & $<10^{-5}$ &$7\times 10^{-5}-0.007$ \\
30 & 6.8& 52 & $> 5 \times 10^4$& $<1.4$& $< 4 \times 10^{-5}$& 0.005 $\pm$ 0.003 \\
31 & 3.14 &52 & $>10^4$ &$<3.2$& $< 4 \times 10^{-5}$ & 0.011 $\pm$ 0.008 \\
32& 2.53 &150 &$> 2 \times 10^4$& $<2.9$ & $< 2\times 10^{-5}$ & $10^{-5}-0.02$\\
\hline
\end{tabular}
\end{table*}

Table \ref{seds} shows that all the sources are older than $10^4$ years.
 Similarly, the cYSOs have low/intermediate masses, lower than 5  $M_{\odot}$. 
The envelope accretion rates are  low, in agreement with the estimated ages \citep{rob07}. In fact, those sources (\#13, 23, and 24) with ages on the order of $10^6$ years are not anymore accreting mass from their envelopes.

As mentioned in Section \ref{co}, Sh2-187 is a young \hii\, region of about $2 \times 10^5$ years, which suggests that the star responsible for the ionized gas might have been born together with the cYSOs, but given its higher mass, it evolved faster and is already in the main sequence phase.

 In summary, given all the observational evidence, we may conclude that 
$i$) \g\, was probably created by a SN explosion, and $ii$) 
a young \hii\, region (Sh2-187) and several low mass cYSOs are located within its boundary.

In this scenario, and bearing in mind that shocks in expanding shells are widely 
believed to be the primary mechanism for triggering star formation \citep{elm98},
it seems possible that the expansion of \g\,  has led to the formation 
of a new generation of stars; as has already been proposed for other shell-like structures 
\citep[e.g.][]{pat98, oey05,arn07, cic09, sua12}.

 The best way to test this hypothesis would be to compare the age of the different objects. 
Concerning the age of \g\,,  assuming that it was created by a SN explosion, it is only about $0.7 \times 10^6$ years (see Section \ref{hi}), which it is not old  enough to generate the formation of all those stars listed in Table \ref{seds}. 
However, the action of the massive progenitor of the SNe should also be
considered. In this sense, the molecular gas where the new stars are being formed
was previously affected by the stellar winds of the star that already exploded as
a SN. If, for illustrative purposes, we consider an O7 star as the  progenitor of
the SN, its lifetime while in the main-sequence is about  $6.4  \times 10^6$ years
\citep{sch92}, which is more compatible with the observed cYSOS. Thus,
a plausible scenario is that the shocks of both the massive progenitor and the
SN explosion could be responsible for triggering star formation.

\section{Summary}

In this work we analyzed a new molecular shell, named \g\,, detected at ($l, b$) = (126\fdg1, --0\fdg8) in the velocity range between --10.5 and --15.5 \kms\, with the aim of elucidating its origin as well as its possible role in triggering the formation of a new generation of stars. 
Our results are summarized as follows:

\begin{enumerate}

\item The  radial velocity of the molecular shell \g\, coincides with the H109$\alpha$ radial velocity estimated for the \hii\, region Sh2-187, indicating that both features are likely to be at the same distance (1.4 kpc). Based on the molecular data, for \g\, we have estimated  an effective radius of about 13 pc,  an expansion velocity of 4.5 \kms, a shell molecular mass of about  $6.5 \times 10^4$ M$_{\odot}$, a mechanical energy of the order of $1.3 \times 10^{49}$ erg, and a  dynamical age  in the range (0.7--1.7)  $\times 10^6$ years.

\item The \hi\, emission distribution presents a cavity of low emissivity having a good morphological correlation with the molecular shell. An estimation of the atomic mass bordering the cavity gives an \hi\, mass of about 300 M$_{\odot}$.

\item Using radio continuum data at 408 and 1420 MHz, we investigated the nature of the diffuse emission detected in the area of \g. The analysis of the TT-plots suggests a non-thermal origin for the emission.

\item An inspection of the mid-infrared emission distribution shows the absence of a photo-dissociated region related to \g.

\item The stellar content located inside \g\, is not capable to inject the mechanical energy needed to create the shell.

\item Both the dynamical age and the required energy of the shell are consistent with the shell being the relic of a SNR. The presence of the pulsar B0105+65 having a similar age and distance as \g\, reinforces this possibility. 

\item Using the available infrared point source catalogues we found several cYSOs projected onto the molecular gas. 
An analysis of their estimated parameters  reveals that they  are low mass stars, most of them are still accreting mass, and have ages greater than  $10^4$  years. 

\end{enumerate}

 Based on all the presented observational evidence, we conclude that \g\, is likely the result of a supernova explosion that took place about $10^6$ years ago.  We also conclude that the expansion of the shell has very likely triggered the formation of several stars, being most of them  low mass stars  still in the accretion phase, while the most massive has already reached the main-sequence phase and is currently ionizing Sh2-187.

\section*{Acknowledgments}
The  DRAO Synthesis Telescope is operated as a national facility by the
National Research Council of Canada.
The CGPS is a Canadian project with
international partners and is supported by grants from NSERC.
Data from the CGPS
is publicly available through the facilities of the Canadian
Astronomy Data Centre (http://cadc.hia.nrc.ca) operated by the 
Herzberg Institute of Astrophysics, NRC. 
This research has made use of the SIMBAD database and VizieR catalogue access tool, CDS, Strasbourg, France. The original description of the VizieR service was published in A\&AS 143, 23.
This publication makes use of data products from the Wide-field Infrared Survey Explorer, which is a joint project of the University of California, Los Angeles, and the Jet Propulsion Laboratory/California Institute of Technology, funded by the National Aeronautics and Space Administration.
This work was partially financed  by the Consejo Nacional de Investigaciones Cient\'{\i}ficas y T\'ecnicas (CONICET) of Argentina under project PIP 112-200801-01299 and Universidad Nacional de La Plata (UNLP) under project 11/G091.
S.C. acknowledges Dr. A. Chernomoretz for his very useful help in the  data analysis. 
\label{lastpage}

\bibliographystyle{mn2e}  
\bibliography{G126-revised}

\IfFileExists{\jobname.bbl}{}
{\typeout{}
\typeout{****************************************************}
\typeout{****************************************************}
\typeout{** Please run "bibtex \jobname" to optain}
\typeout{** the bibliography and then re-run LaTeX}
\typeout{** twice to fix the references!}
\typeout{****************************************************}
\typeout{****************************************************}
\typeout{}
}

\end{document}